\documentstyle[aps,prl,multicol,psfig]{revtex}

\draft

\begin{document}

\title{Knowing a network by walking on it: emergence of scaling}

\author{Alexei V\'azquez$^{1,2}$}

\address{$^1$ Abdus Salam International Center for Theoretical Physics\\
        Strada Costiera 11, P.O. Box 586, 34100 Trieste, Italy}

\address{$^2$ Department of Theoretical Physics, Havana University\\
        San L\'azaro y L, Havana 10400, Cuba}

\maketitle

\begin{abstract}

A model for growing networks is introduced, having as a main ingredient that new
nodes are attached to the network through one existing node and then explore the
network through the links of the visited nodes. From exact calculations of two
limiting cases and numerical simulations the phase diagram of the model is obtained. 
In the stationary limit, large network sizes, a phase transition from a network with
finite average connectivity to a network with a power law distribution of
connectivities, with no finite average, is found. Results are compared with
measurements on real networks. 

\end{abstract}


\begin{multicols}{2}\narrowtext

A network is composed by a set of nodes and a set of links among then. The
topological properties of disordered networks have been studied for a long time. A
well known example is the work of Erdos and R\'enyi \cite{bolobas85} where a network
generated by placing links among the nodes at random is studied. However such a
model is not able to describe the topological properties of real complex networks.
The main current studies are thus focused in finding the mechanism which generates
such networks
\cite{watts98,nunes00,solomon00,jain98,christensen98,slalina99,huberman99,barabasi99}. 

Watts and Strogatz \cite{watts98} introduced the "small-world" network model, an
interpolation between regular lattices and random graphs. Different social,
biological and economic networks has been found to be well described by such
approach \cite{watts98,nunes00}. Such a model is more appropriate for networks where
the number of nodes remains constant. Moreover, it yields a distribution of
connectivities $P(k)$ peaked around a characteristic value \cite{barabasi99}. 

On the other hand, some authors have studied the topological properties of networks
generated by evolutionary dynamics \cite{christensen98,slalina99}. In this case the
network topology is changed using extremal dynamics rules inspired in the
Bak-Sneppen model for biological evolution \cite{bak-sneppen}. A model with fixed
\cite{christensen98} and variable \cite{slalina99} number of nodes has been
proposed. The study of the topological properties of the second model reveals that
the distribution of connectivities changes in time, yielding either exponential or
power law distributions. 

Finally, there is a class of growing-network models where the addition of new nodes
leads to scale-free structures \cite{huberman99,barabasi99}.  In this case the
connectivity distribution follows a power law decay $P(k)\sim k^{-\gamma}$
($2\leq\gamma\leq3$).  Examples are the World Wide Web (WWW) where HTML documents
are the nodes and the links to other documents in the WWW are the links
\cite{barabasi99,broder00};  and the citation of scientific publications where
papers are the node and citations among them are the links \cite{redner98}. In these
two examples the number of nodes (HTML documents or papers) is clearly increasing in
time. Here the attention is focused in this class of networks. 

Different points of view appear when describing the evolution in time of the set of
links, which is actually the mechanism introducing randomness in scale-free models.
In the approach by Huberman and Adamic \cite{huberman99} the number of links
pointing to a node is a random fraction of the number of links which are already
pointing to that node. On the other hand, Barab\'asi and Albert \cite{barabasi99}
have proposed a preferential attachment, where new nodes are linked with higher
probability to those existing nodes with have higher connectivity. This model has
been recently shown to be a particular case of a model proposed by Simon in the
fifties \cite{bornholdt00}. The study of this class of growing networks is currently
very active and new variants have been proposed
\cite{dorogotsev00,barabasi00,krapivsky00}, keeping the preferential attachment as
main ingredient. 

However, these growing-network models do not take into account one fundamental
property of real networks, the fact that a new node does not have "knowledge" of the
entire network. For instance, when a scientist is writing a manuscript he does not
know all the already published papers which may have certain relation with the
subject he is dealing with. In fact he only knows a few number of papers and through
the references appearing on them he found new ones, and continues his search
recursively using the new references on them. Thus, the model introduced in this
paper is based on the fact that we know the network, or at least part of it, by
"walking" on it. This feature together with growing yield the scaling behavior
observed in real growing networks. 

The model is defined by giving an initial condition and a set of evolution rules. 
{\em Initial condition}: one starts with one node $N=1$ and an empty set of links. 
The evolution rules are divided in {\em adding} a node and {\em walking} through the
network. {\em Adding}:  A new node $N+1$ is created with a link to one existing node
selected at random. {\em Walking}: the new node "walks" through all the nodes
pointed by the selected node and create a link to them with a probability $p$. This
last rule is repeated recursively with the new selected links. When no new link is
created add a new node.

\begin{figure}
\centerline{\psfig{file=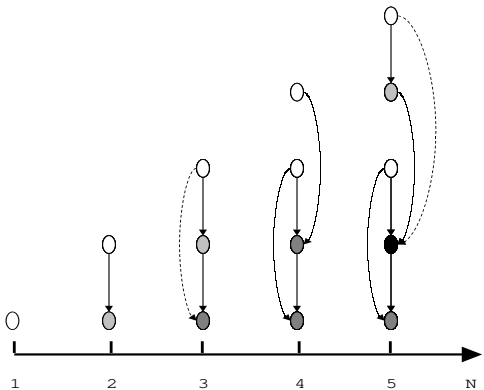,width=2.5in,height=2in,angle=0}}
\caption{One run up to 5 nodes of the algorithm with generates the network using
$p=0.5$. The number of nodes in the network is indicated in the horizontal axis.
Different gray levels indicate different connectivities from $k=0$ (white) to $k=3$
(black).  Dashes lines indicates that the new node performed one "walk" before
creating this link.}
\label{fig:1}
\end{figure}

One run of this algorithm for $p=0.5$ and up to $N=5$ nodes is shown in Fig. 
\ref{fig:1}. $N=1$: a node is created (node 1). $N=2$: a second node (node 2) is
created which can only point to node 1. $N=3$: node 3 is created and it can point
either to node 1 or 2. In the particular case shown in Fig. \ref{fig:1} it
points to node 1. Since node 1 does not has any link the rule stops.  $N=4$: node 4
is created which can point to either node 1,2 and 3. In this case
it points to node 2. Now node 2 has a link to node 1 so with probability $p$ node 4
creates a link to node 1 (it is not created in this case). $N=5$:
node 5 is created which can point to either node 1,2,3 and 4. In this case
it points to node 4. Since $4$ has a link to node 2 node 5 will create a
link to node 2 with probability $p$ (it is created in this case).
But now node 2 has a link to node 1 so with probability $p$ node 5 creates a link to
node 1 (it is not created in this case). And so on. 

The main assumptions of this model is that one has the first contact with the
network through one node and then explores the rest of by "walking"  through the
directed links. Moreover, there is a time scale separation between the addition of
nodes and the mechanism of creation of new links. The network is clearly a directed
graph and between two nodes there can be only one link, which goes from one to the
other. The only parameter of the model is $p$ which may have different
interpretations according to the particular problem one is modeling. For instance,
in the problem of citations $p$ is the fraction of papers appearing in the list of
references of one paper which may be of our interest.

Let us now investigate the evolution of the connectivity distribution as $N$ grows. 
Here the connectivity is defined as the number of links pointing to a node
(in-degree). When a new node is added to the network the connectivity of any node
already at the network remains constant or increases by one. For instance, in Fig.
\ref{fig:1}, from $N=3$ to 4 the connectivity of node 2 increases by one while that
of the other nodes remain constant. Moreover, the created node has connectivity
$k=0$

Let $w(k,N)$ be the probability that when adding the $N+1$ node the connectivity of
a node with connectivity $k$ increases by one. With this definition, the number of
nodes $n(k,N)$ with connectivity $k$ evolves according to the set of equations
\begin{equation}
n(0,N+1)=n(0,N)+1-w(0,N)n(0,N),
\label{eq:1}
\end{equation}
\begin{eqnarray}
n(k,N+1)=n(k,N)+w(k-1,N)n(k-1,N)-
\nonumber
\\
-w(k,N)n(k,N),\ \ \ \ \text{for}\  k>0. 
\label{eq:2}
\end{eqnarray}

For $N\gg1$ one can look for stationary solutions of this set equations. In
this limit $w(k,N)$ should be of the form $w(k,N)=W(k)/N$, where $1/N$ comes from
the fact that the new node is attached to an existing node selected at random, which
happens with probability $1/N$ (this will be demonstrated below for two
limiting cases). Then, taking into account that $n(k,N)=NP(k,N)$ and the stationary
condition $P(k,N+1)=P(k,N)=P(k)$ one obtains
\begin{equation}
P(0)=W(0)/2,.
\label{eq:4}
\end{equation}
\begin{equation}
P(k)=W(k-1)/[1+W(k)],\ \ \ \ \text{for}\ k>0.
\label{eq:3}
\end{equation}
Thus, determining $W(k)$ one can iterate (\ref{eq:3}) to obtain the stationary
distribution $P(k)$. 

A node with connectivity $k=0$ can only increase its connectivity if the new node is
attached to it, which happens with probability $1/N$. Hence, $w(0,N)=1/N$ and,
therefore, $W(0)=1$. From this result and (\ref{eq:4}) it follows that $P(0)=1/2$.
This result is independent of the value of $p$, i.e. in the present model half of
the nodes have no links pointing to them.

The form of $w(k,N)$ for $k>0$ is not known. Here only the limiting cases $p=0$ and
$p=1$ are solved exactly. For $p=0$, independent of the connectivity of a node, the
probability that its connectivity increases by one is $w(k,N)=1/N$, which is just
the probability that the new node is attached to it. Hence, $W(k)=1$ independent of
$k$. Substituting this result in (\ref{eq:3}) and iterating with the initial
condition $P(0)=1/2$ it results that
\begin{equation}
P(k)=2^{-(k+1)},\ \ \ \ \text{for}\ p=0.
\label{eq:6}
\end{equation}
In the other limit, $p=1$, a node will increase its connectivity either if the new
node is attached to it or to one of the nodes with a link to it, i.e.
$w(k,N)=(1+k)/N$. In this case after iteration of (\ref{eq:3}) one obtains
\begin{equation}
P(k)=[(k+1)(k+2)]^{-1},\ \ \ \ \text{for}\ p=1.
\label{eq:7}
\end{equation}

Notice that for $p=1$ since $w(k,N)=(1+k)/N$ there is a preferential attachment to
nodes with larger connectivity. This is one of the main ingredients introduced by
Barab\'asi and Albert to obtain the desired emergence of scaling \cite{barabasi99}.
Actually in their model $w(k,N)$ is also linear in $k/N$. However, the preferential
attachment is imposed while here, on the contrary, it appears self-consistently from
the dynamics of the network. The evolution rules of the model do not show us {\em a
priori} the existence of a preferential attachment but it is clear that nodes with
larger connectivity becomes more visible when one "walks"  through the network. 

These limiting cases are described by distributions which are qualitative different. 
For $p=0$ the distribution is exponential with a finite average connectivity. On the
contrary for $p=1$ the distribution is the power law decay $P(k)\sim k^{-2}$ for
large $k$. This power law decay goes up to the largest possible connectivity $k=N-1$
while $P(k,N)=0$ for $k\geq N$. Then for large $N$ the average connectivity scale as
\begin{equation}
\langle k\rangle=A+\ln N,
\label{eq:8}
\end{equation}
where $A$ is independent of $N$. The average connectivity thus diverges when
$N\rightarrow\infty$.

Since the limiting cases $p=0$ and $p=1$ give qualitative different behaviors there
should be certain probability threshold $p_c$ where a transition from a network with
finite average connectivity to a free-scale network takes place. In the absence of
analytical results for $0<p<1$ numerical simulations are performed in order to
explore this part of the phase diagram. 

\begin{figure}
\centerline{\psfig{file=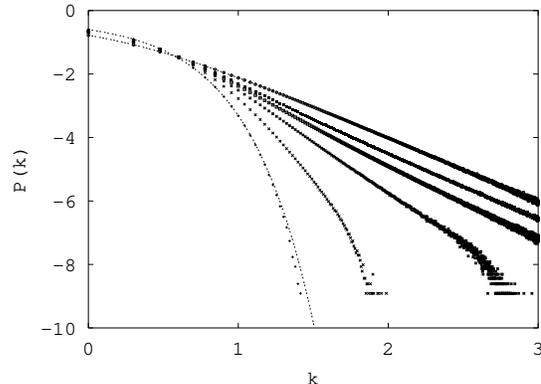,width=3in,angle=-90}}
\caption{Connectivity distribution for different values of $p$. The points
corresponds to, from left to right, $p=0$, $p=0.1$, $p=0.2$, $p=0.3$, $p=0.4$,
and $p=1$, respectively. The continuous lines corresponds with the exact
results for $p=0$ and $p=1$ in Eqs. (\ref{eq:6})  and (\ref{eq:7}), respectively.}
\label{fig:2}
\end{figure}

The maximum network size reached was 81920 nodes and average was taken over 100 runs
of the algorithm which generates the network, for each value of $p$ reported here. 
The resulting connectivity distribution is shown in Fig. \ref{fig:2}. The first
thing to be notice is that the analytical results in (\ref{eq:6}) and (\ref{eq:7})
for the limiting cases $p=0$ and $p=1$, respectively, are in very good agreement
with the numerical data. Second, the transition from a finite average distribution
to a power law takes place at an intermediate probability $0<p_c<1$, where $p_c$ is
in the neighborhood of $0.4$. 

\begin{figure}
\centerline{\psfig{file=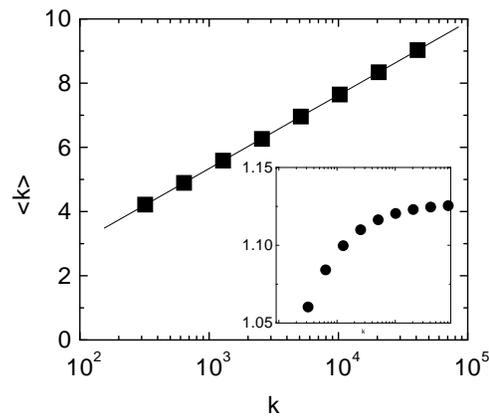,width=2.5in,angle=0}}
\caption{Average connectivity as a function of the number of nodes $N$ added to the
network for $p=0.1$ (inset) and $p=0.9$ (full plot), in a semi-log scale. For
$p=0.1$ the plot clearly saturates to a finite value. On the contrary, for $p=0.9$
$\langle k\rangle$ grows logarithmically, which is manifested as a straight line in
the semi-log plot.}
\label{fig:3}
\end{figure}

To obtain a more precise estimate of the threshold the scaling of the average
connectivity $\langle k\rangle$ with $N$ was investigated. For $p<p_c$ it was found
to saturate to a finite value when $N\gg1$ while for $p>p_c$ it grows
logarithmically with $N$ as in the limiting case $p=1$. The results for $p=0.1$ and
$p=0.9$ are shown in Fig. \ref{fig:3}. In the neighborhood of the threshold the
network sizes needed to reach the stationary state are not accessible by the present
numerical results. Thus, the following approximate method was used to determine
$p_c$. 

The numerical data was fitted by the parabola $\langle k\rangle(N)=a+bx+cx^2$,
where $x=\log N$ and $a$, $b$ and $c$ are fitting parameters. For $p>p_c$ the
parameter $c$ is expected to be zero while for $p<p_c$ it is negative, as a
consequence of the tendency of $\langle k\rangle(N)$ to saturates to the stationary
value. Actually due to numerical errors $c$ will never be exactly zero. Here $p_c$
is estimated by the value of $p$ at which $c$ changes sign, becoming either zero or
positive. Using this criteria it results that $p_c=0.39\pm0.01$. 

Let us now focus our attention in the form of $P(k)$. In Fig. \ref{fig:2} it can
be seen that for $p<p_c$ the shape of the connectivity distribution depends on $p$. 
For $p>p_c$, although there are some deviations for $k$ small, the large $k$
behavior is characterized by the power law decay $P(k)\sim k^{-\gamma}$ with an
exponent $\gamma=2.0\pm0.1$. Thus, above the threshold, the connectivity
distribution of the network is very robust, showing little variations when $p$
changes.

These features are very similar to those observed in some sandpile models
\cite{tadic97,vazquez00}, the paradigm of the theory of self-organized criticality
\cite{bak87}. As in these models, there is a time scale separation, here between the
addition of new nodes and their "walk" through the network. In the thermodynamic
limit, large system sizes, the phase diagram of the model is divided in a
sub-critical and a critical region, and in the critical region the power law exponent
does not depend on the control parameter. All these similarities put these models in
the same class, with a self-organized critical region in the phase diagram. 

There are real complex networks which can be described by the present model. The
network of citations among papers published in journals is an example. In
this case $k$ is the number of times a paper is cited in other papers. The analysis
of the available data yield the power law exponent $\gamma=3$ \cite{redner98}. This
value is larger that the universal value $\gamma=2$ observed in the critical region
of the model introduced here. Thus, it seems that citation problem is in some part
of the sub-critical region below $p_c$. Actually the fraction of papers one usually
considers, to be referred in a future publication, is small in comparison with the
total amount of referenced papers appearing in papers of our knowledge. However,
more data is needed to reach to a final conclusion.

In the WWW network HTML documents are the nodes, the links to other documents in the
WWW are the links, and $k$ is the number of times a HTML document appears as a link
in other HTML documents. In HTML documents the links are created when the HTML
document is created but can also be changed latter on.  Hence, the addition of new
nodes is not the only mechanism of changing the set of links. However, the rate of
addition of new HTML documents is actually very high so one expect that the addition
of new documents is the dominant mechanism and, therefore, can be described by the
present model. Measurements reported in the literature \cite{barabasi99,broder00}
yield the exponent $\gamma=2.1\pm0.1$ in very good agreement with the exponent
$\gamma=2$ in the critical region. Hence, the WWW network is in some part of 
the critical region.

Thus, with only one control parameter the present model is able to describe the
form of the connectivity distribution of networks with different topologies. For
$0\leq p<p_c$ it describes networks with a finite average connectivity, which may
have a power law decay for small connectivities but with a cutoff independent of the
network size for large connectivities. On the contrary, for $p_c<p\leq1$ it
describes networks with power law distribution of connectivities, up to a cutoff
determined by the network size. The transition from one behavior to the other is
determined by the parameter $p$, which measures the probability to create a new link
and continue the search in the network through the added link.

\section*{Acknowledgements}

I thanks S. Redner and L. A. Nunes-Amaral for helpful comments and references.

\vskip 0.1in

\end{multicols}

\end{document}